\documentclass[runningheads]{cl2emult}

\usepackage{makeidx}  
\usepackage{graphicx} 
\usepackage{cropmark} 
\usepackage{eso}      
\makeindex            

\begin{document}
\title*{Mining the Sky with Redshift Surveys }
\toctitle{Mining the Sky with Redshift Surveys}
%
%
\titlerunning{Mining the Sky with Redshift Surveys }
%
\author{Marc Davis \inst{1} \& Jeffrey A. Newman \inst{1}}

\authorrunning{M. Davis and J. Newman}
%
%
\institute{University of California, Berkeley, CA 94720, USA}

\maketitle              

\begin{abstract}
Since the late 1970's, redshift surveys have been vital for progress
in understanding large-scale structure in the Universe.  The original
CfA redshift survey collected spectra of 20-30 galaxies per clear
night on a 1.5 meter telescope; over a two year period the project
added $\approx 2000$ new redshifts to the literature.  Subsequent
low-$z$ redshift surveys have been up to an order of magnitude larger,
and ongoing surveys will yield a similar improvement over the
generation preceding them.  Full sky redshift surveys have a special
role to play as predictors of cosmological flows, and deep pencil beam
surveys have provided fundamental constraints on the evolution of
properties of galaxies.  With the 2DF redshift survey and the SDSS
survey, our knowledge of the statistical clustering of low-redshift
galaxies will achieve unprecedented precision.  Measurements of
clustering in the distant Universe are more limited at present, but
will become much better in this decade as the VLT/VIRMOS and
Keck/DEIMOS projects produce results.  As in so many other fields,
progress in large scale structure studies, both observational and
theoretical, has been made possible by improvements in technologies,
especially computing. This review briefly highlights twenty years of
progress in this evolving discipline and describes a few novel
cosmological tests that will be attempted with the Keck/DEIMOS survey.

\end{abstract}

\section{The Original Center for Astrophysics Survey}

Up until the late 1970's, most extragalactic spectroscopy was dependent on 
photographic plates, or photographic plates in contact with image intensifier
tubes.  Vidicon TV tubes had too little dynamic range and were too unstable for
reliable spectroscopy, and devices such as the Robinson-Wampler scanner 
\cite{wamp75}
had low efficiency and were not true photon counting detectors. 
Needless to say, one did not undertake redshift surveys with 
such tools.

A breakthrough in technology was Steve Shectman's ``Shectograph''
\cite{shect78}, which consisted of a self-scanned Reticon diode array optically
 contacted to a chain of
image intensifier tubes, with a novel analog-digital signal chain for 
counting arriving photon flashes once only and sending addresses to a computer 
interface.  At the Harvard-Smithsonian Center for Astrophysics, minicomputers
such as Data General Nova-2 machines were for the first time being built
into a new telescope, the MMT, and a group of us made use of the available 
engineering talent  to copy 
a Shectograph and interface it to a spare Nova-2 machine with
16 Kbyte of memory and a 7'' diameter floppy disk with 128 Kbyte 
capacity as a mass
storage medium.  Everything was programmed in Forth, an incredibly compact
language that treated the entire computer memory as a push-down stack.  It was
all great fun, but Forth programs had a tendency to be ``write only'', unintelligible after they were created.

\begin{figure}
\centering
\includegraphics[width=\textwidth]{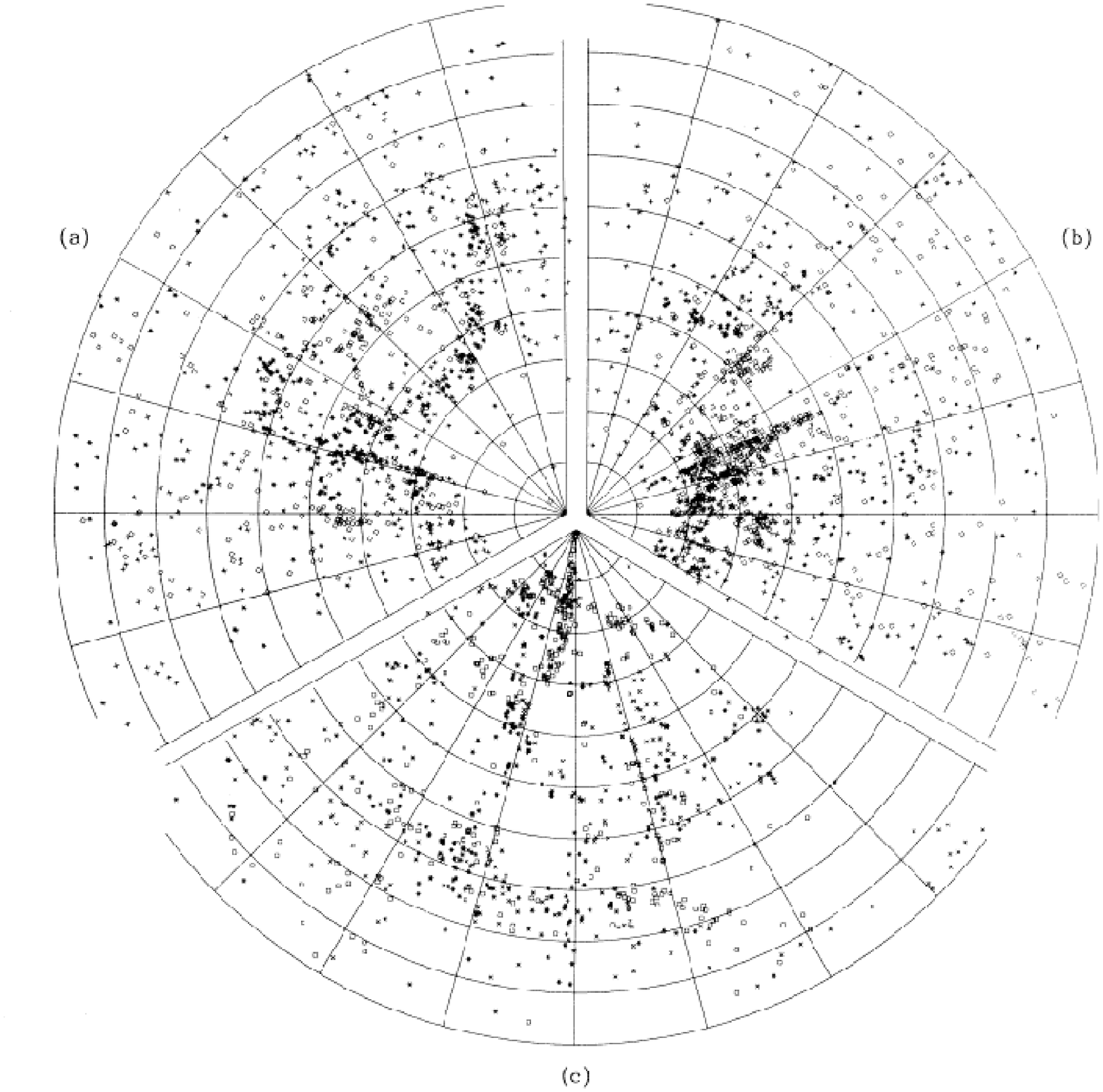}
\caption[]{Redshift space diagrams of the CfA1 survey and two mock CDM 
simulations \cite{defw85} drawn from an $\Omega_m=0.2$ open cosmology.  The 
radial shells are spaced every 1000 km/s}
\label{eps1}
\end{figure}

The Shectograph coupled to a rebuilt low dispersion spectrograph was
immediately dubbed the ``z-machine'', and was installed on the 1.5m Tillinghast
telescope (the one with a spherical primary mirror) on Mt. Hopkins, Arizona.
In those days our competition for telescope time was nil, and the entire
dark time for a two year period was dedicated to execution of the Center
for Astrophysics Redshift Survey, whose first results were reported in 1981
\cite{davis81}.  All data reduction for the project, including the 
cross-correlation redshift analysis \cite{tonry79} 
was done on another Nova-2 computer in Forth.

Once the data was collected, 
we saw for the first time the filamentary nature of
the large-scale structure (LSS) of the Universe, which bore no resemblance whatever
to the predictions of models of the time.  But the Cold Dark Matter revolution
\cite{peebles82} 
was soon to follow, and the infamous ``Gang of Four''  produced mock
universes that were not a bad representation of what was observed in the real
Universe \cite{defw85}.  
At the {\it Mining the Sky} conference, 
I asked the audience to identify the real 
Universe in figure 1, and the response was random (proving once again that
nobody actually reads the literature).  The reader will have to guess or
look up the original reference to learn which of the three wedge plots of figure
1 corresponds to the actual CfA1 survey.

\section{Additional Optically-Selected Redshift Surveys of the Nearby Universe}

From the CfA1 survey and the alphabet soup of 
those that followed, the SSRS surveys 1 and 2, the 
CfA2 survey, the LCRS, APM, CfA Century Survey, the ESO Key slice Survey, 
the ORS, the
ESO neaby Abell Cluster Survey, the ESP,
 the Durham UKST Survey, the CFRS, LDSS, CNOC1,
and CNOC2 surveys, we learned about the 
robustness of $\xi(r)$  and $P(k)$ 
in  real and redshift space, we found that the first CDM models predicted too
little power on large scale and had to be superceded by models with
$\Gamma = .2-.3$,  we learned how 
sensitive the measure of the pair velocity dispersion $\sigma_{12}(r)$ 
was to sample size and to the presence of rich clusters of galaxies, and 
we gained some information
on higher order correlations, especially from the largest surveys such as
the LCRS project.  The LCRS survey has also demonstrated the outer scale of
clustering, (or in  Kirshner's phrase ``the end of greatness'')
with no structures having size comparable to the survey dimensions.
daCosta \cite{dacosta99} provides a more detailed review of past and current
surveys and
tabulates redshift survey parameters.

We heard at this conference the current status of the two major ongoing
redshift surveys, the 2dFGRS  \cite{colless97,colless00} and the SDSS \cite{gunn95}.
These surveys are now both in high gear, each 
collecting more than two CfA1 surveys worth of
galaxies per night.  Clearly, we have come a long way in the past 20 years!
CCDs plus increasingly powerful and inexpensive computers have made all this
wonderful activity possible and have totally transformed astronomical research.
With samples approaching $10^6$ galaxies, we will be able to address
more detailed questions about the nature of the large scale galaxy 
distribution, the nature of bias in the galaxy distribution, and many other
issues.  Indeed, these surveys are central to the approaching era of 
precision cosmology.

\section{Full Sky Redshift Surveys}

Optically selected catalogs of galaxies are of course biased by the extinction 
from dust in the Milky Way, and thus all optically selected redshift 
surveys must be limited to zones of relatively high galactic latitude.  The
Infrared Astronomy Satellite (IRAS) was NASA's first cryogenic satellite,
launched in 1983.  IRAS imaged the full sky in four channels ($12\mu$, $25\mu$,
$60\mu$, and $100\mu$)
and led to the first catalog of galaxies
unaffected by extinction, including approximately 15,000 galaxies
with flux $f_{60\mu} >0.6$ Jy, selected over roughly 85\% of the sky.

Even before any redshifts of IRAS galaxies were available, it was noted that 
the dipole anisotropy
 of IRAS-selected galaxies was aligned reasonably
well with the direction of motion of the Local Group of galaxies as inferred
from the dipole anisotropy of the CMBR \cite{meiksin88,yahil88}.  
This is to be 
expected if the Local Group motion is caused by inhomogeneity in the local
mass distribution and if the IRAS galaxies at least approximately trace the
large scale distribution of mass. 

Several full-sky redshift surveys were undertaken
in the late 1980's and 90's to pursue this lead and to clarify the
scale of the source of CMBR dipole anisotropy.  The first two surveys
QDOT \cite{mrr90}, and the 1.9Jy survey \cite{strauss91}, contained approximately 2500 objects, but the
later surveys, the 1.2Jy survey \cite{fisher94} and the PSCz survey 
\cite{saunders98} contained 5400 and 13,000 galaxies respectively.
Naturally the latter surveys are
superior for statistical inference, but all the IRAS subsamples yield
quite consistent results.  A number of lessons were learned from the
IRAS surveys:

\begin{itemize}

\item
The-full sky galaxy distribution is highly anisotropic up to a redshift of
4000 km/s, but beyond that distance it becomes increasingly isotropic, as
expected in any cosmological model based on the Robertson-Walker metric.  
The cumulative gravitational dipole anisotropy (assuming IRAS galaxies trace the
mass) grows steadily out to 4000 km/s, and then levels off; approximately 2/3 of
the acceleration of the Local Group relative to the comoving frame seems
to arise from within this radius.  The misalignment of the CMBR dipole 
direction and the IRAS acceleration is less than $15^\circ$ in the PSCz
\cite{schmoldt99}, lending strong support to the notion that IRAS galaxies
do trace the mass distribution on large scales.  Nonlinear velocity-field
effects plus
shot noise in the discrete galaxy distribution are sufficient to account for the
remaining misalignment of the dipole vectors.

\item
In the limit of linear perturbation theory, the observed peculiar velocity 
field of nearby galaxies, as observed by such means as Tully-Fisher studies,
is expected to be aligned to the local gravity field, which can be traced 
approximately by the IRAS galaxy distribution.  Although there has been
considerable historical controversy over the mismatch between the gravity
and velocity fields \cite{strauss94,davis96}, 
recent Tully-Fisher samples such as the
Shellflow project \cite{willick99} and the Cornell sample \cite{dale99} show
remarkable consistency with the predictions of the IRAS flow maps. In the
local frame of reference, the peculiar velocities of a 
shell of galaxies at 6000 km/s exhibit a reflex dipole anisotropy of amplitude
$\approx 600$ km/s in a direction opposite to the CMBR dipole.  Thus 
this shell's velocity relative to  the comoving frame of reference 
is quite low, as expected from the 
large scale power spectrum of favored models of structure formation.

\item
The consistency of the velocity and gravity fields allows one to solve
for $\beta \equiv \Omega_m^{0.6}/b$, 
where $b$ is the bias of the IRAS galaxies.
Current best estimates derived from these studies is $\beta = 0.5 \pm 0.1$
\cite{saunders99}
consistent with $\Omega_m = 0.3$ if IRAS galaxies trace the mass distribution. 
 But we know that IRAS galaxies are less strongly clustered than optically
selected galaxies; if the optical galaxies trace the mass distribution,
the large-scale velocity field studies imply $\Omega_m = 0.2$.  Given the
strength of the clustering of the local galaxy distribution, one would
expect the Local Group velocity to exceed 1000 km/s if $\Omega_m = 1$, instead
of the observed $627 \pm 22$ km/s derived from the COBE DMR data 
\cite{smoot91}.

\item
The IRAS galaxy distribution provides an amusing test of the statistical
 homogeneity of the Universe.  In a fractal galaxy distribution, the 
correlation length of the galaxy clustering, $r_0$, should scale linearly
with the radius of the survey.  A test of volume-limited subsamples
of the 1.2Jy IRAS survey \cite{davis94} instead shows that $r_0$ is unchanged
as the limiting volume changes by a factor of 8.  This, plus the
clear approach to isotropy at large scales in the IRAS galaxy distribution,
 would seem to rule out a fractal model, but alas, 
idealogues may never be convinced by data.

\end{itemize}
\subsection{New Full-Sky IR Selected Surveys}

The DENIS and 2MASS projects have now been completed and facilitate the
construction of new, nearly full-sky catalogs of galaxies flux-limited in
the $K$ band.  Two surveys are in advanced stages of development and will
soon begin.  The 6dF survey \cite{colless00a} is to use a 
new 150-fiber spectrograph on the UK Schmidt Telescope in Australia.  The
sample selection will be drawn from 2MASS and DENIS objects limited to 
$K<13$, with an expected median redshift $z \approx 0.05$.  Their target
is to survey 90,000 galaxies in the 17,000 square degrees in the Southern
sky, plus an additional 35,000 targets complete to $H=13.4, J=14.1, I=15.0$,
and $B=16.5$.  The total number of new redshifts will be approximately $10^5$.
This instrument should be commissioned around the end of 2000 -- early
2001.  Huchra \cite{huchra00} 
is leading the 2MASS redshift survey, whose goal is to observe
150,000 galaxies to $K=12.2$ over the full sky.  A second phase of this 
project will be observe $10^5$ galaxies to $K_s=13.5$ using 1:10 sampling.

A major goal of these surveys will be to extend predictions of large-scale
flow field to greater depth than has been possible with the IRAS-selected
samples.  Furthermore, it is known that IRAS galaxies are mostly dusty
spirals undergoing star formation, and therefore undercount dense regions 
in which early type galaxies predominate.  $K$ band flux, on the other hand, is
a fair measure of the stellar mass of a galaxy, and if the mass-to-light
ratios of galaxies are reasonably constant, it should be possible to construct
a more precise gravity field map than with the IRAS samples, for which one
must give each source equal weight.  

\section{Existing Faint Galaxy Redshift Surveys}

All the surveys described thus far are designed to characterize the properties
of galaxies and their distribution at the current epoch, $z \approx 0$.  Even
the 2dFGRS and SDSS will be limited to $z<0.3$, and so are not intended to 
describe the {\it evolution} of galaxies and their clustering properties.
The evolutionary aspects of galaxies and LSS can only be studied by 
analysis of faint, distant samples with large telescopes. As these
 facilities are expensive and in high demand, redshift surveys of the 
distant Universe
have to date been quite limited in extent.  The best constraints thus far
on the evolution of LSS derive from angular clustering studies of faint
galaxy samples \cite{postman98},  but this should change in the next few
years.

The CFRS  \cite{lilly95,lilly97} was the path breaking redshift
survey to probe both
the properties of distant galaxies and their large-scale clustering
statistics.  The survey studied 591 galaxies to a limit $I_{AB} < 22.5$
over an area of 0.14 square degrees, with a median redshift $z\sim 0.6$.  
The survey was undertaken on the CFH telescope, with $15\AA$ resolution
and allowing multiple galaxies per row on the CCD to increase multiplexing.  The CFRS provided unprecedented
constraints on the evolution of the luminosity distribution function and 
suggested that the clustering length of galaxies diminishes quite rapidly with
redshift.  However, the survey volume was very small, and clustering analyses within
less than fair sample volumes have a tendency to be biased low.

The CNOC2 survey \cite{yee98} was a somewhat shallower 
field galaxy survey of 6500 galaxies to a limiting
magnitude of $R<21.5$ or $B<22.5$ over 1.5 square degrees of sky.  The median
redshift of this sample was $z\sim 0.35$, and galaxies ranged from $0.1<z<0.7$.
This project was also executed on the CFH telescope with resolution and methods similar to CFRS.  The survey was executed in four separate
fields, each of which was elongated on the sky to increase the cross-scan
dimensions of the survey.  This survey design lends itself to more
reliable measurements of the galaxy clustering, with results summarized by
Carlberg \cite{carlberg99}.  One of their key results is that the comoving
correlation length of galaxies decreases extremely slowly in the interval
$0.1<z<0.6$, more slowly than the mass correlation length even in low
density cosmologies.  This supports the notion that the bias of the galaxy 
distribution is increasing with redshift, as expected in hierarchical models
\cite{frenk88}

The Caltech Faint Galaxy Survey \cite{cohen99} includes 1200 galaxies with
$22<R<23.5$.  This survey was executed with the LRIS spectrograph on the Keck
telescope, with $\sim5 \AA$ resolution.  The median redshift of this
sample is $z\sim 0.55$, with a tail extending beyond $z=1$.  The sample
contains prominent sheets of galaxies at selected redshifts, and is probably
too small for definitive constraints on the evolution of structure, but
the sample has been a goldmine for study of the evolution of properties of
galaxies, including merger rates, star formation rates, and luminosity functions.  A recent summary is given by Cohen \cite{cohen00}.

\section{Future Faint Galaxy Redshift Surveys}
Two large projects to study the distant galaxy distribution 
are now in their final construction phase.  The VLT/VIRMOS project 
\cite{lefevre00} is on track to begin observations in June, 2001, 
while our team using the Keck telescope and DEIMOS spectrograph plans to begin observations in early 2002 
\cite{davis00}. 
Each of these massive machines build on past experience and contain multiple
CCD detectors for more extensive multiplexing.  Each project plans an
observing campaign in excess of 100 nights on the world's largest telescopes.

\subsection{The VLT/VIRMOS Project}

The VLT/VIRMOS project \cite{lefevre00} is a complex, multi-thrust survey
with several components.  VIMOS is a 4-barreled instrument that is
capable of simultaneous spectroscopy of 800 galaxies with resolution
200.  As with the CFH surveys, the slitmasks will be milled to
place the spectra of multiple galaxies on one row of the CCD detectors. 
Higher resolution spectroscopy will be possible with 
lower multiplexing. The wide-field
component of VIMOS will gather spectra of $10^5$ galaxies to a limiting
magnitude of $I_{AB}=22.5$, the same as for the CFRS, but now over 9 square
degrees.  The medium deep survey will reach to $I_{AB}=24$ for 50,000 galaxies
in a field of 1.2 square degrees. Both of these surveys will be 
undertaken at low resolution.   Higher resolution spectra will be obtained for $\approx 10000$ galaxies.

\subsection{The Keck/DEIMOS Redshift Survey}
The DEIMOS spectrograph is nearing completion in the UC Lick shops and will
be delivered to the Keck Observatory in 2001.  The instrument is the most
complex ever built for the Keck Observatory and as such has presented 
numerous challenges.  To get a feel for the scale of the beast, DEIMOS
has a 1.5m parabolic collimator mirror, the same diameter as the 
primary mirror in the telescope used for the original CfA redshift survey!
Instead of the 8Kbyte data files of the z-machine, each output
from DEIMOS will be 140Mbytes in size.  The most remarkable statement of the
progress in technology 
is that the current computing challenge is not particularly
more difficult than that faced 20 years ago.

 Once DEIMOS is operational, a team of 
astronomers will initiate a major redshift survey of 
galaxies that will consume approximately 120 Keck nights over a 
three year period.  The goal is to gather high quality spectra on 
$\approx 60,000$ galaxies  with $z>0.7$ 
in order to study the evolution of the properties 
 and large scale clustering of galaxies at $z \approx 1$.  
The survey will be done at high
spectral resolution, $R=\lambda/\Delta \lambda \approx 3500$, to work 
between the bright OH sky emission lines and to infer linewidths for
many of the target galaxies.  Many of the science projects planned for this
survey depend on linewidth information, which cannot be derived with
low resolution spectra.  Thus the Keck/DEIMOS survey and the VLT/VIMOS project
will be complementary rather than simply competitive.


The Keck/DEIMOS survey  is a collaborative project between astronomers at 
UC, Caltech, and the
Univ of Hawaii, in addition to outside collaborators.  Team members with
Keck access are M. Davis, S. Faber, D. Koo, R. Guhathakurta, C. Steidel, 
R. Ellis, J. Luppino, and N. Kaiser.

\section{Details of the Keck/DEIMOS survey}

We anticipate that the
workhorse grating for the Keck/DEIMOS survey will be the 900 lines/mm grating, 
which
will provide a spectral coverage of 3500 $\AA$ in one setting.  If we use
slits of width 0.75", they will project to a size of 4.6 pixels, or
a wavelength interval of 2 $\AA$.  Thus the resolving power of the observations
will be quite high, $R \equiv \lambda / \Delta\lambda =3700$.
The spectrum of each galaxy will be dispersed across two CCDs, for a total
of 8k pixels.
The spatial direction of DEIMOS covers 16' of sky, 8k pixels long over 4 CCDs.

The MIT-LL CCDs used in DEIMOS have exceptionally low readout noise, 1-2 $e^-$, and the time
to become sky-noise limited is less than 10 minutes, 
even when using a 1200 l/mm grating.  The large number
of pixels in the dispersion direction allows high resolution with
substantial spectral range,  so that we can work between  the bright OH
sky emission lines while remaining sky-noise limited.

We will set the grating tilt so that the region 6300-9300 $\AA$ is centered on
the detector, thus assuring that the 3727 $\AA$ [OII] doublet is in range
for galaxies with $0.7 < z < 1.5$.  At the planned spectral resolution, 
the velocity resolution will be 80 km/s.  
The [OII] doublet will thus be resolved for all luminous galaxies,
giving confidence to the redshift determination even if no other features are 
observed.  With sufficient flux it should be possible to measure the velocity 
broadening of the lines, which will hopefully lead to an estimate of the
gravitational potential-well depth of a substantial fraction of the 
galaxies within the survey.



\subsection{Fields and Photometry}
\begin{table}
\caption{Fields Selected for the Keck/DEIMOS Redshift Survey}
\begin{center}
\begin{tabular}{|c|c|c|c|} \hline
RA & dec & (epoch 2000) & mask pattern \\ \hline
14$^h$ 17 & +52$^\circ$ 30  &  Groth Survey Strip  & 120x1 \\ \hline
16$^h$ 52 & +34$^\circ$ 55   & zone of low extinction & 60x2\\ \hline
23$^h$ 30 & +0$^\circ$ 00   & on deep SDSS strip & 60x2 \\ \hline
02$^h$ 30 & +0$^\circ$ 00   & on deep SDSS strip & 60x2 \\ \hline
\end{tabular}
\end{center}
\end{table}

The Keck/DEIMOS survey will be undertaken
in four fields, as listed in Table 1.  The fields were chosen as low
extinction zones that each are continuously observable at favorable zenith 
angle from Hawaii over a six month interval. 
One field includes the Groth
Survey strip, which has good HST imaging and which will be the target of very 
deep IR imaging by SIRTF, and two of the fields are on the
equatorial strip that will be deeply surveyed by the SDSS
project.  Each of these fields is the target of a CFHT 
imaging survey
by Luppino and Kaiser, whose primary goal is very deep imaging for weak 
lensing studies.  They are using the new UH camera (8k by 12k pixels) with
a field of view of 30' by 40', in the B, R, and I bands.
The imaging will be obtained in random pointings over
a field of $3^\circ$ by $3^\circ$, but with continuous coverage of a strip
of length $2^\circ$ by $30'$ in the center of each field.

Given this enormous photometric database, we shall use the color information
 to exclude galaxies with $z<0.7$.  DEIMOS will be used
to undertake a spectroscopic survey of the remaining galaxies which have $m_I(AB) \le 23.5$.  At this relatively bright flux limit, 60\% of the
galaxies should be at $z<0.7$ \cite{lilly97,cohen99}; the photometric
redshift preselection eliminates this foreground subsample, 
allowing us project to focus our effort on the high-redshift Universe.
Although the depth of the
Keck/DEIMOS sample will be comparable to the deep VLT/VIMOS
survey, it will contain a somewhat larger number of high redshift objects.

\subsection{Observing Strategy--Target Selection }

In each of the four selected fields of Table 1, we plan to densely target a
region of 120' by 16' or 120' by 30' for DEIMOS spectroscopy. 
We intend to produce
120 separate masks per field; each mask 
will each contain slitlets selected from  
a region of size 16' by 4', 
with the slitlets mostly aligned 
along the long axis, but with some tilted as much as $30^\circ$ to track
extended galaxies.  Our goal is to select an average of 
 130 slitlets per mask, selected from the
list of galaxies 
meeting our flux and color cuts.  The mean surface
density of candidate galaxies  exceeds the number of objects we
can select  by approximately 30\%, 
and spectra of selected targets cannot be allowed to overlap.  However, 
this will not cause problems with the subsequent analysis if we take account
of the positions of those galaxies for which we did not obtain spectroscopy.
In the Groth strip region, because of the interests of other 
collaborative scientific
projects such as SIRTF, the plan is to construct 120  masks each
covering a sky patch offset from its neighbors 
by 1', and to select targets without regard to color.  Thus any spot
on the sky will be found within 4 masks, giving every galaxy
 4 chances to appear
on a mask without conflict. In the other three survey fields, we plan to
to use the color preselection, halving the source density of targets, and
to step 2' between masks, giving a galaxy two chances to be selected without
conflict.  In these fields the masks will form a pattern of 60 by 2, covering
a field of 120' by 30'.  At $z \approx 1$, this field subtends a comoving
interval of $80~\times~20 h^{-2}$ Mpc$^2$, and our redshift range translates to
a comoving interval of $\approx 800h^{-1}$ Mpc (in a flat Universe with 
$\Omega_\Lambda =0.7$).



\subsection{Two Surveys in One}  
Our team's project is actually subdivided into two surveys, 
the 1HS (one hour integration
survey) and the 3HS (three hour integrations).  The 1HS is the backbone 
project and will require 90 nights on Keck for execution.  The planned
large scale structure studies are the main drivers of the 1HS design.  The 
3HS will require 30 nights of Keck time to obtain spectra of $\approx 5000$ 
targets. The major
scientific focus of the 3HS is to study 
the properties of galaxies, but it will also provide a critical check on
the quality of linewidths and other properties derived from 1HS 
spectra.  3HS targets will be selected to a
flux limit up to one magnitude fainter than the 1HS.
The 3HS  survey fields will be limited to a few 16' by 4'
regions, plus one  field 16' by 16' in the center of the Groth Survey zone,
which will have very deep SIRTF imaging.
We intend to acquire HST imaging in all 3HS fields, as well as Chandra
and XMM images.


\subsection{Keeping up with the Data}

The data rate from DEIMOS will be in excess of 1 Gbyte/hour, so 
automated  reduction and analysis tools are absolutely imperative.  We
have been working closely with the SDSS team and intend to adapt the
IDL code of Schlegel and Burles for our pipeline reduction.  
The photometric and
spectroscopic databases for the project are currently planned to be
IDL structures,
stored on disk as FITS binary tables.  Although the total raw data will exceed
one Tbyte, the reduced data will be modest in size by today's standards,
$< 50$ Gbytes.  

The Keck/DEIMOS team intends to  share all results 
with the public and to put the spectra online in a timely manner.  
Details of the project can be found at the URL
http://astro.berkeley.edu/deep/.

\section{Science Goals of the Keck/DEIMOS Survey}

The 1HS survey is designed to provide a fair sample volume for analysis
of LSS statistical behavior, particularly for clustering studies
on scales $ < 10h^{-1} $ Mpc. 
 The comoving volume surveyed in the 1HS program will
exceed that of the LCRS survey \cite{shectman}, a survey 
 which has proven to be an 
outstanding resource for low redshift studies of LSS; the comoving density of galaxies studied will also be comparable to the LCRS. 
When the data is in hand, we plan a number of scientific analyses, with
major programs listed below.

\begin{itemize}
\item
Characterize the linewidths and spectral properties of galaxies versus color,
luminosity, redshift, and other observables. The 3HS is designed
to reach a depth allowing more detailed analyses of the internal properties
of galaxies at high redshift, including their rotation curves, linewidths from 
absorption spectra, and stellar populations.

\item 
Precisely measure the two--point and three--point correlation functions of 
galaxies at $z=1$ as a function of other observables, 
such as color, luminosity,
or linewidth.  For the higher--order correlations,  dense sampling is
essential.  Observations of Lyman-break galaxies at $z \sim 3$ \cite{steidel}
show that the bias in the galaxy distribution was considerably higher in
the past.  Higher order correlations in the galaxy distribution are one way
to estimate the presence of bias in the galaxy distribution \cite{fry}. Furthermore, if the
galaxy bias is larger at $z=1$ than at present, the correlation
strength of different subsamples of galaxies should show more systematic
variation than is observed for galaxies at $z=0$.  The design of the 1HS is
driven by the goal of obtaining a fair sample measure of the two and three
point correlation functions at $z\sim1$ in a volume larger than that of the
LCRS.

\item
Measure redshift space distortions in the galaxy clustering at $z=1$ using the $\xi(r_p,\pi)$ diagram and other  measures.  The
evolution of the thermal velocity dispersion provides another way of separating
the evolution of the galaxy bias from evolution in the 
underlying matter distribution.  The high precision of the
 redshifts provided by 
DEIMOS will make this measurement possible.

\item 
Count galaxies as a function of redshift and linewidth, in order to 
execute the classical redshift-volume cosmological test.  Details are given by
Newman \& Davis \cite{newman}, who show that this test can provide a precision measurement of
the cosmological parameters $\Omega_m$ and $\Omega_\Lambda$. If instead 
one assumes the universe to be flat, the test can set a strong constraint
on the equation of state parameter $w$ of the dark energy component (for which
$P= w\rho$) \cite{steinhardt}.  Examples of the constraints possible from
this test are given in figure 2.  Note the substantial degeneracy between
$w$ and $\Omega_m$; both parameters change the volume element in a
similar fashion.  Additional constraints in this parameter space 
can be used to reduce this 
degeneracy.  An example  results from
analysis of the strength of clustering within the Keck/DEIMOS survey,  as shown
in figure 3, where we plot contours of the ratio of the correlation
integral J3 measured at $z=1$ to that determined locally.  This ratio is a readily
measured quantity which is dependent on the growth rate of structure and thus
sensitive to a different combination of parameters than the volume constraints. The distance between contours equals the estimated 2$\sigma$ error in the measurement of the J3 ratio.  Thus, for 
certain regions of parameter space the two tests together can precisely
constrain {\it both} $w$ and $\Omega_m$.

\item
Separate the Alcock-Paczynski effect \cite{alcock},\cite{ball} from 
the redshift 
space distortions of the $\xi(r_p,\pi)$ diagram.  This effect relates
intervals of angular separation to intervals of redshift separation 
as a function of redshift. 
An object that appears spherical at low redshift would
appear elongated in redshift at $z>0$, but the degree of elongation is
a function of $q_0$.  This effect is 10 times smaller in amplitude than
the redshift-volume effect and will be challenging to measure, but 
it is conceivable that the Keck/DEIMOS project will
provide data that can measure this effect and separate it from the
other expected redshift space distortions.

\end{itemize} 

\begin{figure}
\centering
\includegraphics[width=\textwidth]{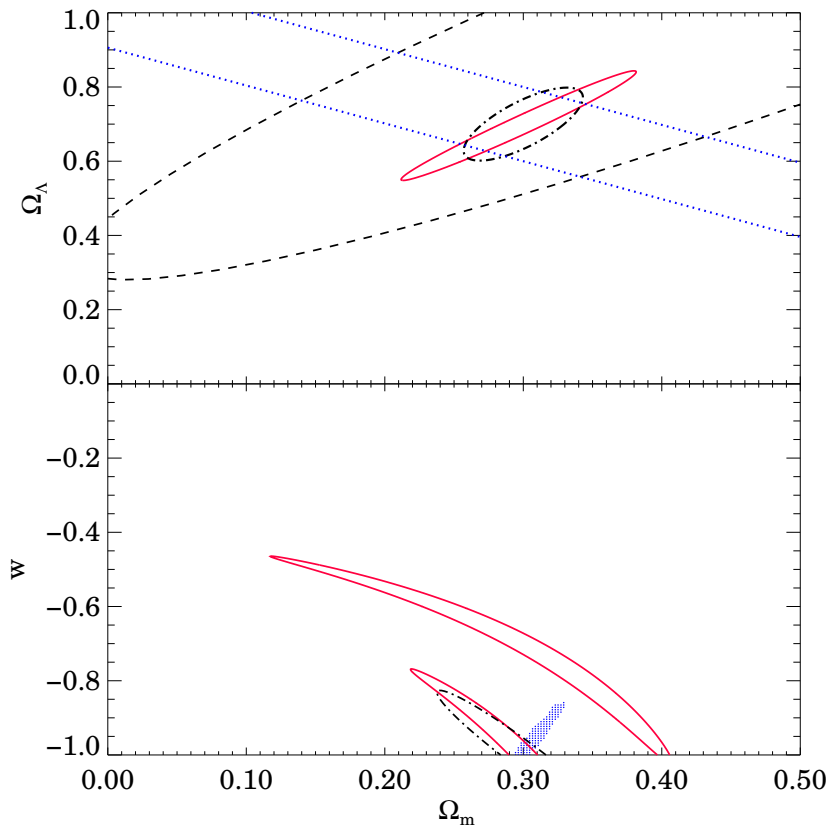}
\caption[]{ Constraints 
in the $\Omega_m$, $\Omega_\Lambda$  or $\Omega_m$, $w$ (bottom) planes for a variety of proposed cosmological tests.  (Top)
The solid curve depicts the 95\% confidence constraint resulting from 
determining the number counts $dN/dz$ using 10,000 galaxies in the interval 
$0.7<z<1.5$ with measured linewidths, as we have proposed for the Keck/DEIMOS 
survey.  The black dashed curve is the 68\% error contour for the recent Type Ia observations of Perlmutter et al., while the dot-dashed curve is the 95\% statistical error contour expected from the proposed SNAP satellite 
\cite{snap}.  The dotted contours emulate constraints resulting from near-term measurements of the position of the first doppler peak in the
CMBR.  The contours for future measurements all presume a $\Omega_m=0.3$, $\Omega_{\Lambda}=0.7$ model.  The $dN/dz$ test can provide constraints comparable in strength to those from SNAP, with more complementarity to CMB measurements. (Bottom) Contours are depicted as above.  The constraints from our proposed measurement of $dN/dz$ are shown for two sample cosmologies, with $\Omega_m$=0.3 and $w=-0.7$ or -1.  The shaded region depicts the constraint which would result from proposed surveys for clusters extending to $z \sim 1$ \cite{haiman}.
}

\label{eps2}
\end{figure}

\begin{figure}
\centering
\includegraphics[width=\textwidth]{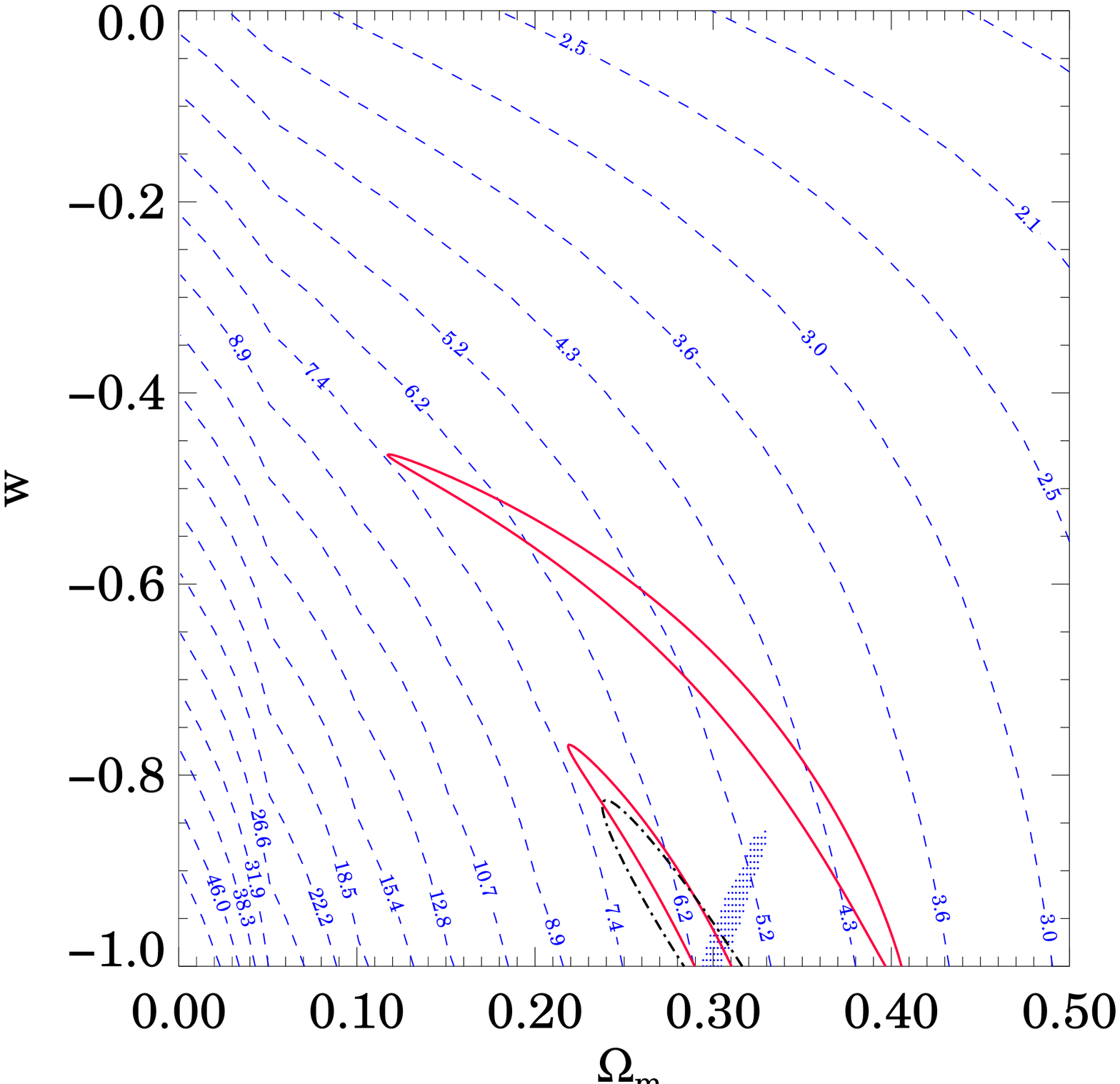}
\caption[]{Constraints in the $w$, $\Omega_m$ plane resulting from
the J3 ratio test.  The dashed curves are contours of constant J3 ratio; other curves are as in the bottom panel of the preceding figure.  The distance between contours is equivalent to the estimated $2\sigma$ error in measuring the J3 ratio from the Keck/DEIMOS survey.  The intersection
of this constraint with the redshift-volume constraint has the potential to
lift the degeneracy between the parameters, independent of other measurements. }
\label{eps3}
\end{figure}

\section{Conclusions}

Over the past two decades, redshift surveys have been central to progress
in the study of large-scale structure.  Different types
of surveys have served diverse scientific goals, and this enterprise is
larger now than ever.  Three distinct classes of survey will be revolutionized in this decade:

\begin{itemize}

\item
The 2dFGRS and SDSS are massive surveys at high galactic latitude which will
provide definitive measures of LSS and galaxy properties in the nearby universe.

\item
The 6dF and 2MASS redshift surveys will provide the best estimate of
the full-sky galaxy distribution in the local Universe.  

\item
The VLT/VIRMOS and Keck/DEIMOS redshift surveys will provide samples of
galaxies at $z\sim1$ with the fidelity of the local LCRS sample, which will
greatly advance our knowledge of the evolution of the properties of galaxies
and of large-scale structure.  

\end{itemize}  

All of these activities will greatly advance our understanding of the
large-scale Universe and provide major clues on questions of fundamental
physics.  Astrophysics has never been so flooded with data, but this
conference has demonstrated that we can build the 
tools needed to mine  these exquisite resources.


\subsection{Acknowledgements}

{This work was supported in part by NSF grants AST00-71048. 
The DEIMOS spectrograph is funded by a grant from CARA
(Keck Observatory), by an NSF Facilities and Infrastructure grant (AST92-2540), and by gifts from Sun Microsystems and the Quantum Corporation.}

\clearpage
\addcontentsline{toc}{section}{Index}
\flushbottom
\printindex

\end{document}